\documentclass[sigconf]{acmart}
\AtBeginDocument{%
  }

\acmYear{2025}\copyrightyear{2025}
\acmConference[NanoCom '25]{International Conference on Nanoscale Computing and Communication}{October 23--25, 2025}{Chengdu, China}
\acmBooktitle{International Conference on Nanoscale Computing and Communication (NanoCom '25), October 23--25, 2025, Chengdu, China}
\acmDOI{10.1145/3760544.3765641}
\acmISBN{979-8-4007-2166-3/25/10}

\usepackage{graphicx}                 
\usepackage{subcaption}               
\begin{document}

\title{Node Position Estimation in Diffusion-Based Molecular Communications Using Multi-Layer Perceptron}
\author{Sangjun Hwang}
\email{sangjun848@yonsei.ac.kr}
\orcid{0009-0009-1323-9647}
\affiliation{%
  \institution{Yonsei University}
  \state{Seoul}
  \country{Republic of Korea}
}

\author{Chan-Byoung Chae}
\email{cbchae@yonsei.ac.kr}
\orcid{0000-0001-9561-3341}
\affiliation{%
  \institution{Yonsei University}
  \city{Seoul}
  \country{Republic of Korea}
}

\renewcommand{\shortauthors}{Sangjun Hwang, Chan-Byoung Chae}

\begin{abstract}
This paper proposes a method for accurately estimating the relative position between two nodes with unknown locations in a diffusion-based molecular communication environment. A specialized node structure is designed, combining a central absorbing receiver with multiple transmitters placed at predefined spherical coordinates. Pilot molecules are released, and their absorption time and concentration are measured. By partitioning the spherical coordinate space, these spatially distinct measurements serve as input to a multilayer perceptron (MLP)-based model. The proposed method significantly improves the precision of distance and direction estimation. Simulation results demonstrate localization accuracy, confirming the effectiveness of the neural network model in capturing the underlying physical characteristics.


\end{abstract}

\ccsdesc[300]{Computer systems organization~Molecular computing}


\keywords{Molecular Communications, Position Estimation, Multi-layer Perceptron}

\maketitle
\section{Introduction and system model}

Introduction: Molecular communication via diffusion (MCvD), which uses chemical signals for nanoscale applications, is a viable alternative to traditional electromagnetic systems~\cite{7405285,chae2023_jcn}. In such applications, a network of cooperative nanomachines must first determine their relative positions to each other before they can route data, fuse measurements, or act in concert. Accurate localization of unknown nodes is crucial for enhancing the reliability and efficiency of such networks.

This work proposes a node structure that combines an absorbing receiver with multiple transmitters. To enable accurate relative position estimation, each node partitions spherical coordinates into octants and transmits pilot molecules from transmitters placed at the coordinate intersections. This study proposes a relative three-dimensional coordinate estimation method using a physical formula and a multi-layer perceptron (MLP). Our approach differs from the existing literature, which primarily focuses on regressing channel parameters from a single transmitter-receiver pair or estimating the coordinates of only a single transmitter point.

System model: Each node is modeled as a spherical absorbing receiver (RX) of radius $r$. The surface of the node is divided into eight octants, which are defined by the polar and azimuthal angle axes. Six transmission tubes (TXs) are mounted at the intersections of these boundaries to release pilot molecules. This multi-transmitter setup allows for repeated distance estimations to improve accuracy and helps infer direction when a path is obstructed by Node A.

\begin{figure}[!t]
    \centerline{\includegraphics[width=8cm]{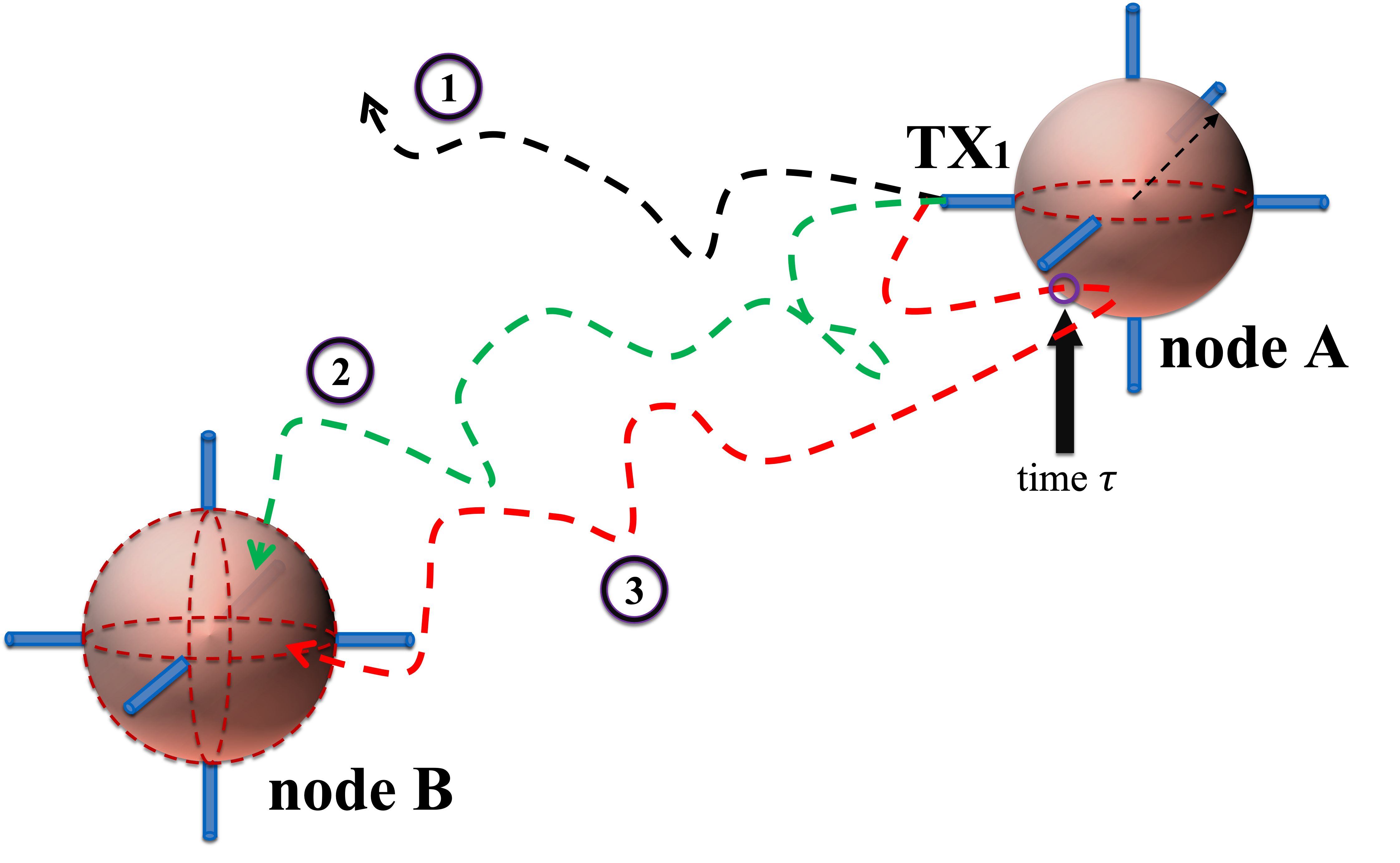}}
\caption{Three scenarios for a molecule released from a transmitter TX1 on Node A. Path~1 is a molecule lost in free space. Path~2 is directly received by Node~B. Path~3 visualizes superposition, where a molecule absorbed at Node~A at time $\tau$ could have probabilistically reached Node~B.}
    \label{figure_1}
    \vspace{-10pt}
\end{figure}

Fig.~\ref{figure_1} describes the system model, illustrating all types of path: the two possible real paths (path 1 and 2) and one imaginary path (path 3)~\cite{kwak2020twoway}.
We consider an MCvD system in which every molecule moves solely by Brownian motion. This system consists of only two nodes, named Node A and Node B, neither of which knows its relative coordinates. The transmitters on Node A emit pilot molecular signals, and Node B estimates the relative location based on this received information. To this end, Node B records the arrival time and the count of absorbed molecules, while Node A simultaneously absorbs a subset of its own emissions.

\section{Position estimation}
Accurate positioning in the estimation process requires both distance and direction information. For a molecule emitted at position $X$ and absorbed at $Y$ with diffusion coefficient $D$, the first hitting time probability density is given by
\begin{equation}
f_{XY}(t) = \frac{r (d_{XY} - r)}{d_{XY} \sqrt{4 \pi t^3 D}} \exp\left[ -\frac{(d_{XY} - r)^2}{4Dt} \right],
\label{fund}
\end{equation}
for time $t$~\cite{yilmaz2014three}.
The calculation also accounts for superposition effect~\cite{ferrari2022channel}. Specifically, if a molecule is absorbed by Node A at time $\tau$, the model also computes the probability of that same molecule reaching Node B in the subsequent time interval $t - \tau$. By applying this principle and assuming sufficiently long guard times between pilot pulses, the expected molecule count at Node B becomes
\begin{equation}
N_B = N \cdot \sum_n \left( \frac{r}{d_{AB}} \right)^{2n} \left( \left( \frac{r}{d_{A_TB}} \right) - \left( \frac{r}{d_{A_TA}} \right) \right),
\label{phys}
\end{equation}
where $N$ is the number of emitted molecules and $A_T$ denotes a transmitter on Node A. 

Within direction information, the set of possible locations for Node A forms a sphere centered on the Node B. Our proposed method first determines a direction of arrival by partitioning the receiver's surface into eight octants and identifying the most likely sector. Subsequently, a machine learning model refines this coarse directional information into a precise position estimate. The input features for the model, computed for each octant, include the peak arrival time, received molecules, and the total count of accumulated molecules.

This estimation process requires pilot signals from at least two transmitters to resolve positional ambiguity. With a single pilot transmitter, TX$_i$, the known range only constrains the possible locations of Node A to a circle centered on the axis connecting TX$_i$ and B. To address this, our implementation performs triangulation by selecting the two pilot signals with the highest received signal strength for both training and inference. Furthermore, we introduce a physically informed loss function, derived from the principles in eq.~\ref{phys}, that incorporates a distance constraint.

\section{Simulation results and concluding remarks }

\begin{figure}[t]
  \centering
  \begin{subfigure}[b]{1\linewidth}
    \centering
    \includegraphics[width=\linewidth]{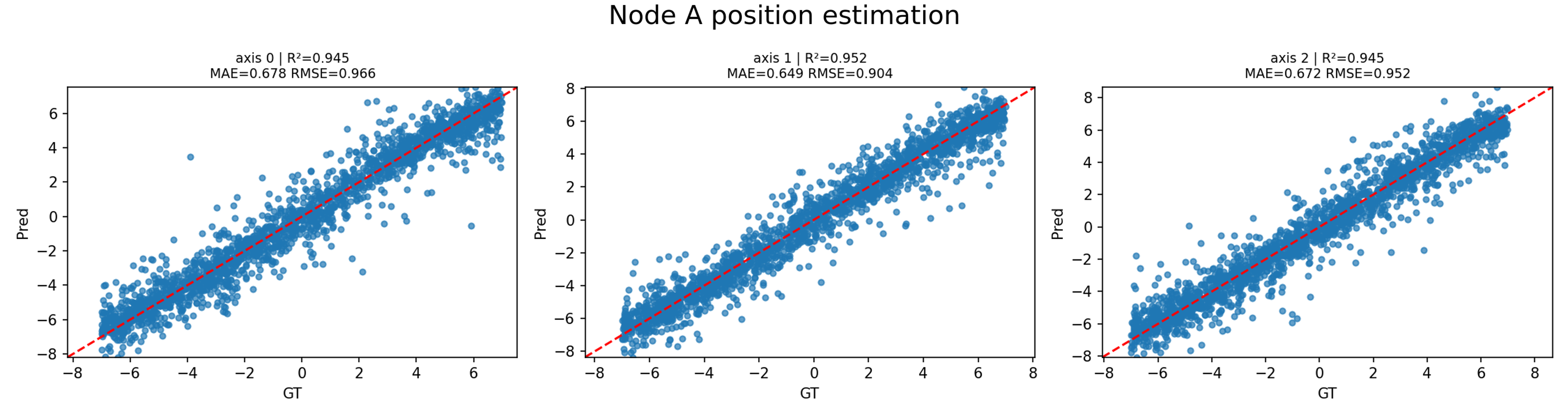}
    \caption{}
    \label{fig:tx_pred}
  \end{subfigure}
  \hfill
  \begin{subfigure}[b]{1\linewidth}
    \centering
    \includegraphics[width=\linewidth]{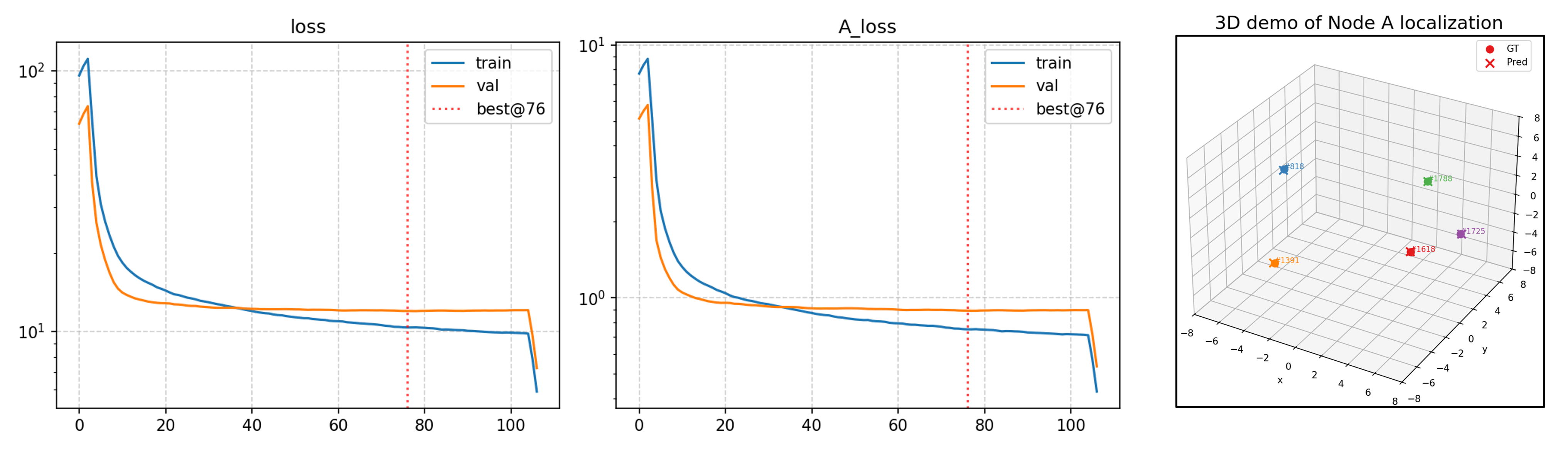}
    \caption{}
    \label{fig:nodeA_pred}
  \end{subfigure}
  \caption{Results for Node A position estimation. (a) Scatter plots of predicted versus ground truth by axis on the test set. (b) The two left panels show the training and validation curves for the total loss and Node A loss, and the right panel illustrates 3D comparisons for five test samples.}
  \label{fig:pred_results}
  \vspace{-5pt}
\end{figure}
Using our molecular communication simulator, we trained an MLP featuring an attention pooling layer. The model was designed to jointly regress the 3D position and orientation of a target node, along with the positions of the TX.

A dataset of 10,000 samples was generated and split for training, validation, and testing, with a standardization scaler fitted only to the training data. As shown in Fig.~\ref{fig:nodeA_pred}, our proposed model achieved a mean $R^2$ of approximately 0.947 for the coordinates of Node A, with a Mean Absolute Error (MAE) of 0.666 and a Root Mean Square Error (RMSE) of 0.941. For the reconstructed transmitter positions, the model achieved an MAE of 0.633 and an RMSE of 0.890.

For comparison, we trained a simple ridge regression model as a baseline on the same data. This baseline achieved a coordinate-wise $R^2\approx$ 0.830, with an MAE of 1.309 and an RMSE of 1.695. Our proposed model significantly outperformed this baseline, reducing the MAE and RMSE by approximately 49$\%$ and 45$\%$, respectively.



In conclusion, we have presented a localization framework that blends analytical diffusion expressions with learned nonlinear mapping has been presented. Our results demonstrate that both transmitter and receiver coordinates can be estimated with high precision. The combination of physics‑informed features and a lightweight neural model therefore offers an effective and computationally attractive solution for positioning in diffusion‑based molecular communication networks. Future research will include identifying the location between nodes through pilot signals in the full-duplex system~\cite{9566506}.

\begin{acks}
This work was supported by National Research Foundation of Korea (NRF) Grant through the MSIT and IITP grants (RS-2023-00208922, RS-2024-00428780, 2022R1A5A1027646).
\end{acks}

\bibliographystyle{ACM-Reference-Format}
\bibliography{refs}

\end{document}